\documentclass[]{spie}  

\usepackage{amsmath,amsfonts,amssymb}
\usepackage{graphicx}
\usepackage[colorlinks=true, allcolors=blue]{hyperref}
\usepackage{threeparttable}
\usepackage[caption=true,font=footnotesize]{subfig}

\title{Deep Convolutional Neural Networks for Molecular Subtyping of Gliomas Using Magnetic Resonance Imaging}

\author[a]{Dong Wei\textsuperscript{*$\dagger$}}
\author[b]{Yiming Li\textsuperscript{*}}
\author[c]{Yinyan Wang\textsuperscript{$\dagger$}}
\author[d]{Tianyi Qian}
\author[a]{Yefeng Zheng}
\affil[a]{X-Lab, Youtu Lab, Tencent, Shenzhen, China}
\affil[b]{Beijing Neurosurgical Institute, Capital Medical University, Beijing, China}
\affil[c]{Dept. of Neurosurgery, Beijing Tiantan Hospital, Capital Medical University, Beijing, China}
\affil[d]{Healthcare Lab, Sinovation Ventures AI Institute}

\authorinfo{\textsuperscript{*} D. Wei and Y. Li contributed equally to this work.\\%
\textsuperscript{$\dagger$} Correspondences: donwei@tencent.com/tiantanyinyan@126.com.}

\begin{document}
\maketitle

\begin{abstract}
\textbf{Purpose: }Knowledge of molecular subtypes of gliomas can provide valuable information for tailored therapies.
This study aimed to investigate the use of deep convolutional neural networks (DCNNs) for noninvasive glioma subtyping with radiological imaging data according to the new taxonomy announced by the World Health Organization in 2016.
\textbf{Methods:} A DCNN model was developed for the prediction of the five glioma subtypes based on a hierarchical classification paradigm.
This model used three parallel, weight-sharing, deep residual-learning networks to process 2.5-dimensional input of trimodal MRI data, including T1-weighted, T1-weighted with contrast enhancement, and T2-weighted images. A data set comprising 1,016 real patients was collected for evaluation of the developed DCNN model.
The predictive performance was evaluated via the area under the curve (AUC) from the receiver operating characteristic analysis.
For comparison, the performance of a radiomics-based approach was also evaluated.
\textbf{Results:} The AUCs of the DCNN model for the four classification tasks in the hierarchical classification paradigm were 0.89, 0.89, 0.85, and 0.66, respectively, as compared to 0.85, 0.75, 0.67, and 0.59 of the radiomics approach.
\textbf{Conclusion}: The results showed that the developed DCNN model can predict glioma subtypes with promising performance, given sufficient, non-ill-balanced training data.
\end{abstract}


\section{INTRODUCTION}
\label{sec:intro}  

Each year, approximately 100,000 people are diagnosed with diffuse gliomas~\cite{CancerStats2018Bray}.
Diffuse glioma is the most aggressive and malignant form of primary brain tumors, which has a high mortality rate~\cite{molinaro2019genetic,jiang2016cgcg}.
According to the most up-to-date taxonomy announced by the World Health Organization (WHO) in 2016, five molecular subtypes of diffuse gliomas are recognized based on isocitrate dehydrogenase (IDH) and 1p/19q genotypes, in addition to histologic phenotypes~\cite{WHO2016louis}.
The knowledge of glioma subtypes is of great value to the planning of effective therapy, yet to obtain this knowledge requires the biopsy process which carries significant risks.
To explore noninvasive alternatives, Lu et al. applied a radiomics approach to predicting glioma subtypes from multimodal MRI, and achieved encouraging results on a relatively small data set comprising 284 patients~\cite{glioma_sub2018Lu}.
Deep convolutional neural networks (DCNNs) have recently gained massive attention due to their success in a wide range of visual tasks, including outstanding performances in numerous medical image applications over traditional image processing and computer vision techniques~\cite{DL_survey2017litjens}.
Few studies have applied DCNNs for partial prediction of glioma subtypes based on multimodal MRI~\cite{yogananda2019novel}.
However, the capability of DCNNs for comprehensive glioma subtyping in accordance with the WHO 2016 taxonomy remains unexplored.
To fill this gap, we present a DCNN model for comprehensive glioma subtyping in this study, and experimentally show that this model can predict glioma subtypes with promising performance superior to that of the radiomics approach.

\section{Materials and Methods}

\subsection{Hierarchical Glioma Subtyping Paradigm}
According to the WHO 2016 taxonomy~\cite{WHO2016louis}, gliomas can be classified using a three-level subtyping paradigm including four binary classification tasks (Table~\ref{tab:three-level}).
At the top level, gliomas are classified into lower grade gliomas (LGGs) and glioblastomas (GBMs) based on histologic phenotypes.
Then at the middle level, both LGGs and GBMs can be classified as with IDH mutation or wild-type IDH (referred to as IDH mut and wt afterwards).
At the bottom level, IDH mut in LGGs are further classified as with 1p/19q codeletion or noncodeletion (referred to as 1p/19q codel and noncodel afterwards).
This generic paradigm for glioma subtyping has been previously used in Ref.~\citenum{glioma_sub2018Lu}.
In this study, we also adopt this paradigm.
Therefore, independent binary classifiers will be built for the four classification tasks.
Compared to a single-level multiclass model, the three-level paradigm is more flexible to plug in histologic and genotype information when available.

\begin{table}[ht]
  \caption{The three-level glioma subtyping paradigm, based on histologic phenotypes, IDH, and 1p/19q genotypes.
  }\label{tab:three-level}
  \begin{center}
  \begin{threeparttable}
    \begin{tabular}{c|c|c|c|c|c|c}
    \hline\hline
    Level & Biomarker &\multicolumn{5}{c}{Glioma}\\
    \hline
    Top & Histology & \multicolumn{3}{c|}{LGG} & \multicolumn{2}{c}{GBM} \\
    \hline
    Middle & IDH & \multicolumn{2}{c|}{Mutation} & Wild type & Mutation & Wild type \\
    \hline
    Bottom & 1p/19q & Codeletion & Noncodeletion &--&--&--\\
    \hline
    \multicolumn{2}{c|}{Subtype\tnote{~a}} & I & II & III & IV & V \\
    \hline\hline
    \end{tabular}
    \begin{tablenotes}\footnotesize
      \item[a] Glioma subtypes: I: 1p/19q codel in IDH mut LGGs, II: 1p/19q noncodel in IDH mut LGGs, III: IDH wt in LGGs, IV: IDH mut in GBMs, and V: IDH wt in GBMs.
    \end{tablenotes}
  \end{threeparttable}
  \end{center}
\end{table}

\subsection{Study Cohort and MRI Data}
\label{sec:title}

\begin{figure}[ht]
    \centerline{
    \hfil
    \subfloat[T1w]{
        \includegraphics[width=.20\textwidth]{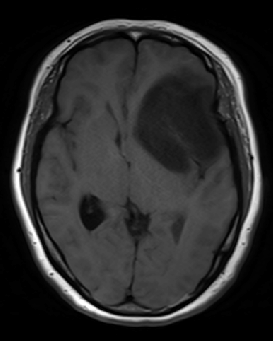}
    }
    \subfloat[T1CE]{
        \includegraphics[width=.20\textwidth]{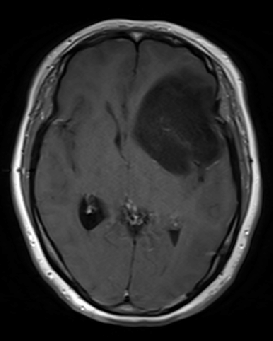}
    }
    \subfloat[T2w]{
        \includegraphics[width=.20\textwidth]{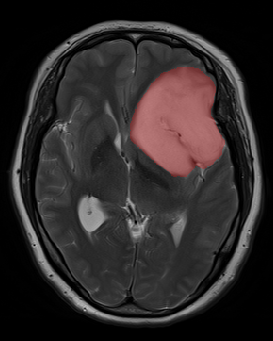}
    }
    \hfil
    }\caption{The preoperative, multiparametric MRI scan for each included patient comprises: (a) T1-weighted (T1w), (b) T1 contrast enhancement (T1CE), and (c) T2-weighted (T2w) sequences.
  The whole tumor region was delineated on the T2w image while referring to the T1w and T1CE images.}\label{fig:data_n_label}
\end{figure}

Ethical approval of this retrospective study was received from the institutional review board of the Beijing Tiantan Hospital, Capital Medical University, Beijing, PR China.
The study cohort was drawn from consecutive patients treated at the hospital from September 2014 to April 2018.
The inclusion criteria were as follows:
(1) pathologically diagnosed as primary diffuse gliomas; (2) available preoperative T1-weighted (T1w), T1 contrast enhancement (T1CE), and T2-weighted (T2w) MRI sequences (Fig. \ref{fig:data_n_label}); (3) age$\geq$18 years; (4) available IDH status (detected with immunohistochemistry or pyrosequencing); and (5) available 1p/19q status (detected using fluorescence in situ hybridization) for LGGs.
In total, 1,016 patients were included.
The 1,016 patients were randomly divided into a training and a test set (concretely, a case was either placed in the training set with 80\% probability, or placed in the test set with 20\% probability, resulting in a 780:236 split), with the former used for model construction and the later for evaluation.
Table~\ref{tab:data} charts the clinical characteristics of the included patients in the training and test sets.

\begin{table}[ht]
\caption{Patient characteristics in the training and test sets.
Please refer to Table~\ref{tab:three-level} for the glioma subtypes.}\label{tab:data}
\begin{center}
\begin{threeparttable}
  \begin{tabular}{lccccccccccccc}
  \hline\hline
  Data & Total & \multicolumn{2}{c}{Age} & & \multicolumn{2}{c}{Sex} & & \multicolumn{5}{c}{Glioma subtypes}\\
  \cline{3-4}\cline{6-7}\cline{9-13}
  set & No. & Median & IQR\textsuperscript{~a} & & Male & Female & & I & II & III & IV & V \\
  \hline
  Train & 780 & 47 & 21 & & 445 & 335 & & 138 & 116 & 191 & 60 & 275 \\
  Test & 236 & 47 & 19 & & 132 & 104 & & 47 & 33 & 47 & 19 & 90 \\
  \hline
  $p$ value & - & \multicolumn{2}{c}{0.349\textsuperscript{~b}} & & \multicolumn{2}{c}{0.819\textsuperscript{~c}} & & \multicolumn{5}{c}{0.617\textsuperscript{~c}} \\
  \hline\hline
  \end{tabular}
  \begin{tablenotes}
  \item[a] Interquartile range.
  \item[b] Mann-Whitney U-test.
  \item[c] Chi-square test.
  \end{tablenotes}
\end{threeparttable}
\end{center}
\end{table}

MRI scans were performed using 3.0T scanners (TrioTim, Siemens: 381 patients; Discovery MR750, GE: 279 patients; Verio, Siemens: 204 patients) and 1.5T scanners (Signa HDe, GE: 152 patients),
including axial T1w images (repetition time, 1750--2459 ms; echo time, 9.4--19.6 ms; slice thickness, 5--7.15  mm),
T1CE images using 0.1 mmol/kg of Gd-DTPA injections (Beijing Beilu Pharmaceutical Co., Beijing China) (repetition time, 1777--2752 ms; echo time, 9.4--19.6 ms; slice thickness, 5--7.15 mm),
and T2w images (repetition time, 4500--7568 ms; echo time, 84--106.3 ms; slice thickness, 5--7.15 mm), with field of view (175--240) mm $\times$ (220--240) mm, and matrix sizes of (192--512) $\times$ (288--512) pixels.
The whole tumor regions were manually outlined by experienced neurologists on the T2w images slice by slice (Fig.~\ref{fig:data_n_label}(c)) using the ITK-snap software (http://www.itksnap.org)~\cite{yushkevich2006user}, while referring to the other two modalities.

\subsection{Image Preprocessing}
Minimal preprocessing is performed.
First, the MRI volumes are resampled to 0.34$\times$0.34$\times$5 mm by cubic interpolation.
Then, the T1w and T1CE volumes are rigidly registered to the T2w volume.
Subsequently, the tumor masks manually delineated on the T2w images are transferred to the T1w and T1CE images.
Lastly, intensities of each MRI volume are normalized to have a zero mean and unit standard deviation.
Note that bias-field correction is not conducted.

\subsection{DCNN Model Development}
\begin{figure}[ht]
 \begin{center}
   \includegraphics[width=.8\textwidth]{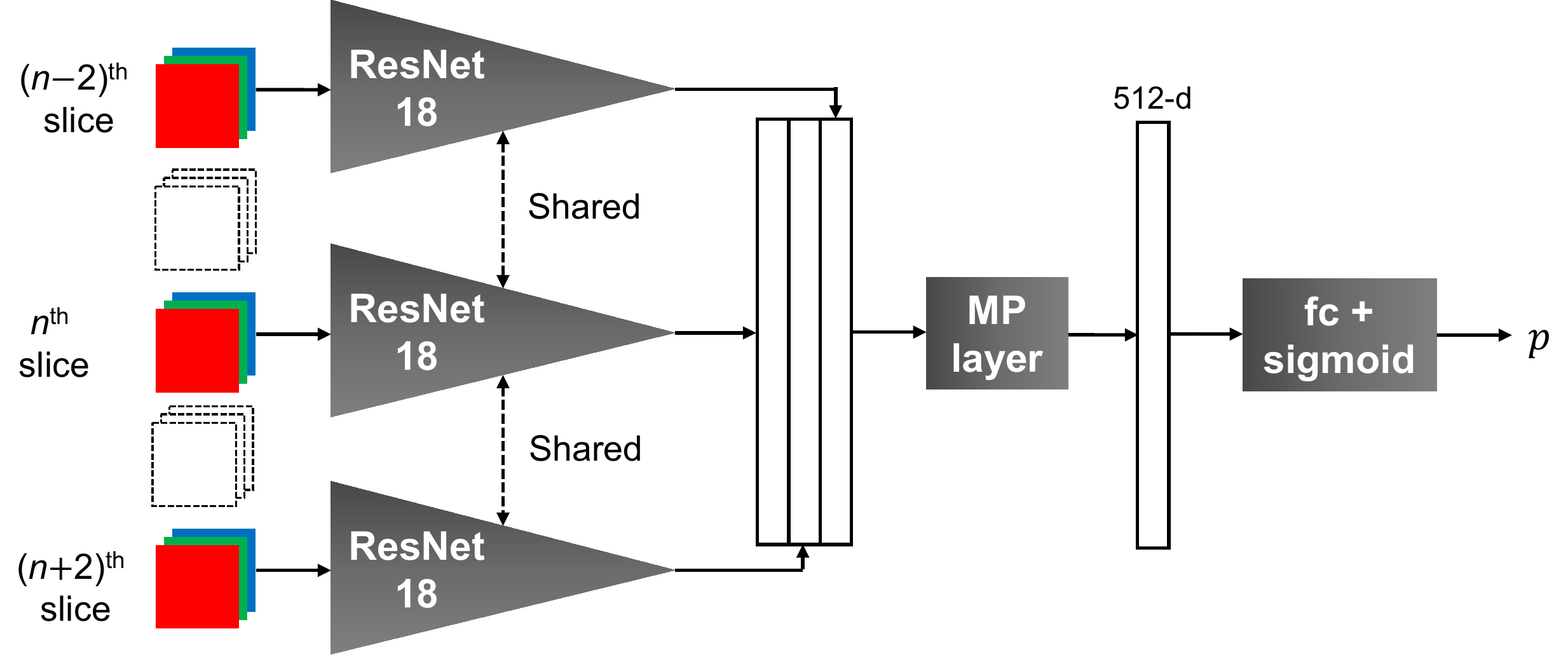}
 \end{center}
 \caption{Network structure of the developed 2.5D DCNN model for glioma subtyping.
 MP = max pooling; fc = fully connected.
 The input to the network are tumor ROIs cropped from three (every other) slices, each of which has three channels (T1w, T1CE, and T2w, represented with distinct colors).
 The output is the probability of being classified as a positive sample.}\label{fig:cnn}
\end{figure}

In this study, we develop a 2.5-dimensional (2.5D) DCNN model (Fig.~\ref{fig:cnn}) for the classification tasks in the glioma subtyping paradigm, based on the residual blocks for deep learning~\cite{resnet2016deep}.
Specifically, we adapt a ResNet18 model pretrained on the ImageNet as the backbone feature extractor, by removing the average pooling and fully connected layers at the end.
The same DCNN model is used for the four classification tasks.
Considering the apparent anisotropy of the data used, 2.5D input is adopted for the DCNN model, instead of 3D.
Specifically, using the tumor mask, the slice with the maximum tumor area is identified;
assuming it is the $n$\textsuperscript{th} slice of an input volume.
Then, the $(n-2)$\textsuperscript{th}, $n$\textsuperscript{th}, and $(n+2)$\textsuperscript{th} slices are extracted and input to the DCNN (see the network illustration in Fig.~\ref{fig:cnn}).
We also experimented with three consecutive slices---i.e., $(n-1)$\textsuperscript{th}, $n$\textsuperscript{th}, and $(n+1)$\textsuperscript{th} slices---instead of every other slice, but found this choice less stable.
The three imaging modalities (T1w, T1CE, and T2w) are treated as the channels of a slice.
A rectangular region of interest (ROI) that can cover the tumor areas in all the three slices is cropped out from each of these slices, and resized to 224$\times$224 pixels.
Next, the resized crops are input to three parallel, identical backbone networks (the triangles in Fig.~\ref{fig:cnn}) with shared parameters, producing  three feature vectors of 512 dimensions (512-d).
Then these three feature vectors are max-pooled (MP) across slices to become a single 512-d feature vector.
This final feature vector is input to a fully connected (fc) layer and a sigmoid function, yielding $p$---the probability of being classified as a positive sample.

To mitigate class imbalance for a given binary classification task (if needed), the small class with substantially less instances is over sampled by integer times to roughly match the number of instances in the bigger class.
Data augmentation is employed to counteract overfitting while training the DCNN models, including:
(1)~the central slice of the 2.5D input is randomly selected from the top three slices with largest tumor areas;
(2) random translation, resizing, left/right mirroring, intensity scaling, and rotation of the ROIs.
No class rebalance or data augmentation is used during validation or testing.

For a given classification task, the training set is randomly divided into five folds, each of which is used as the validation set in turn.
When a specific fold is used for validation, the other four folds are used to train the DCNN model.
The area under the curve (AUC) of the receiver operating characteristic (ROC) curve is used as the stopping and model selection criteria.
Concretely, training is stopped 10 epochs after the training AUC rises above 0.99 for the first time, and the model with the highest validation AUC is selected.
In addition, the optimal operating point is determined on the validation ROC curve where the true positive rate minus the false positive rate is maximal.
Iterating through the five folds results in five individual DCNN models for the current classification task.
At inference time, a test case is fed into all the five models to get five probabilities and five predictions.
The final predictive probability is the mean of the five probabilities, and the final prediction is obtained by a majority voting.

\subsection{Implementation}
The DCNN models are trained on an NVIDIA Tesla P40 GPU using mini-batches of 32 cases.
The binary cross entropy loss is used.
The parameters of the models are optimized using the Adam algorithm~\cite{adam2014kingma} with a weight decay of 0.0005 for L2 regularization.
All parameters of the pretrained layers are fine-tuned during the training process.
The learning rate is initially set to 0.0005, and exponentially decayed by multiplying by the factor $0.97^{Ep}$, with ${Ep}$ being the epoch number.
The DCNNs are implemented using the PyTorch package.

\section{Experimental Results}
\label{sec:exp_n_results}

\begin{figure}[ht]
  \begin{minipage}{\linewidth}
    \centering
    \subfloat[GBM vs. LGG]{
        \includegraphics[width=.35\columnwidth]{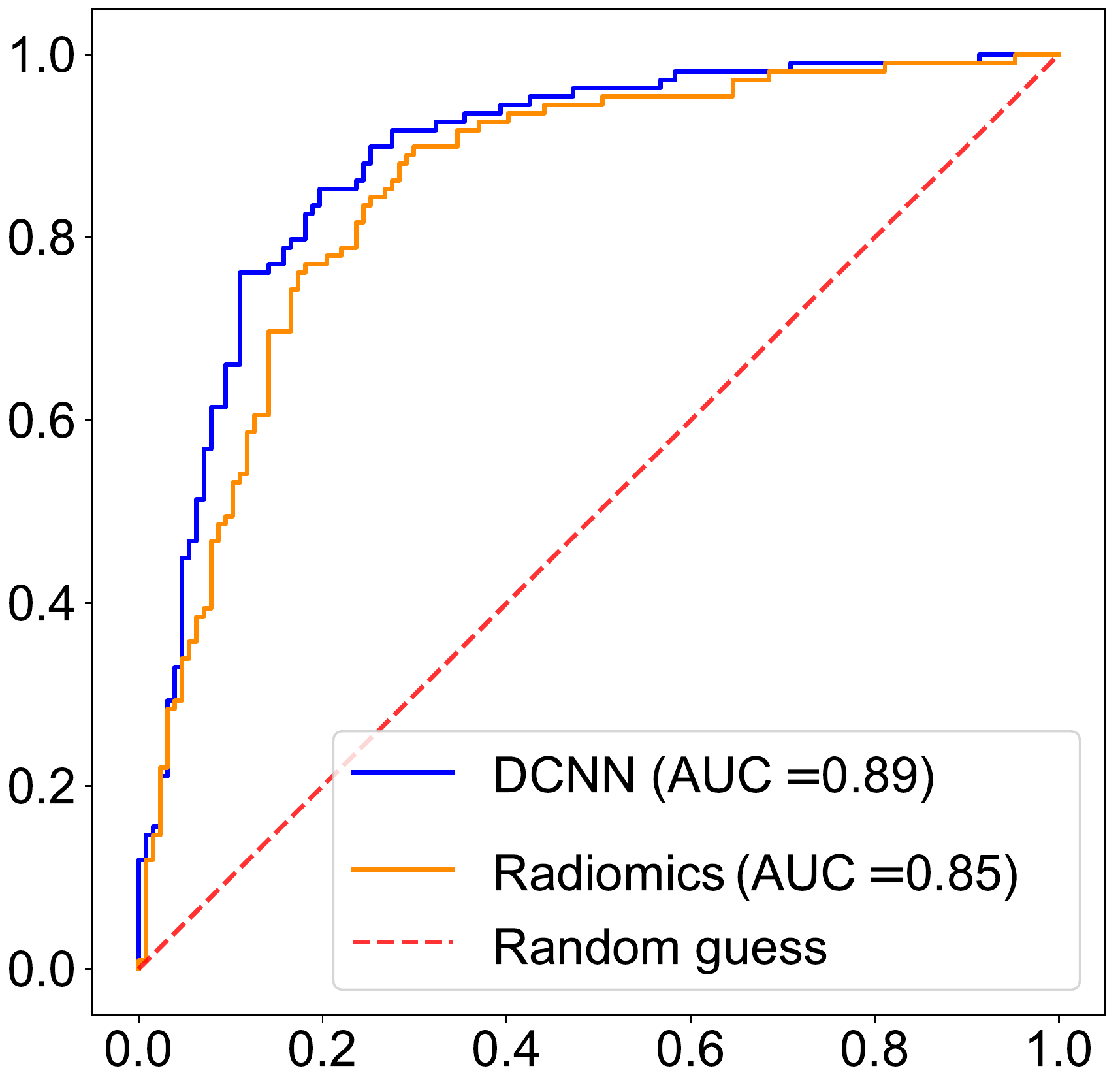}
    }
    \subfloat[IDH mut vs. wt in LGGs]{
        \includegraphics[width=.35\columnwidth]{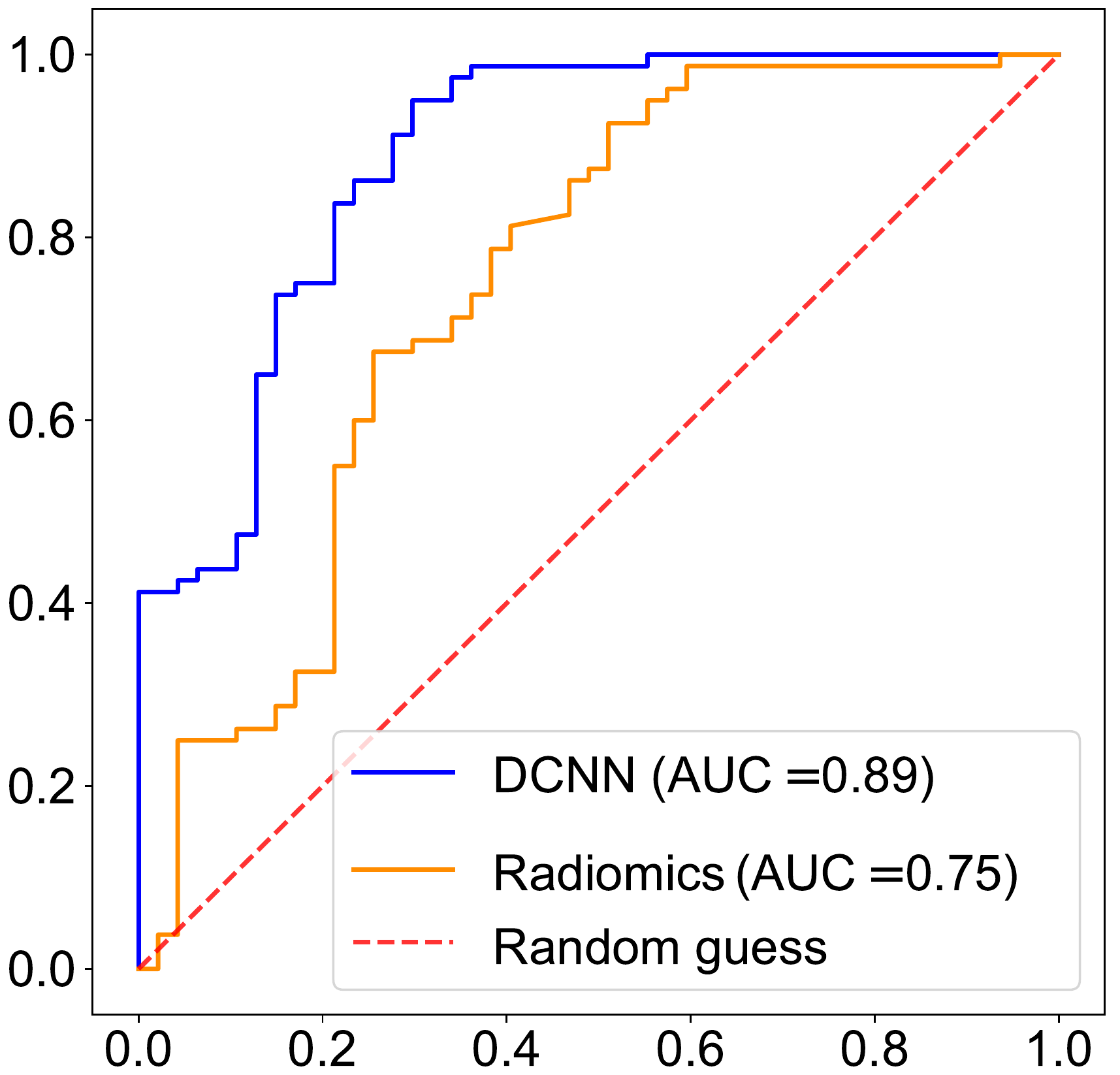}
    }
  \end{minipage}
  \begin{minipage}{\linewidth}
    \centering
    \subfloat[1p/19q codel vs. noncodel in IDH mut LGGs]{
        \includegraphics[width=.35\columnwidth]{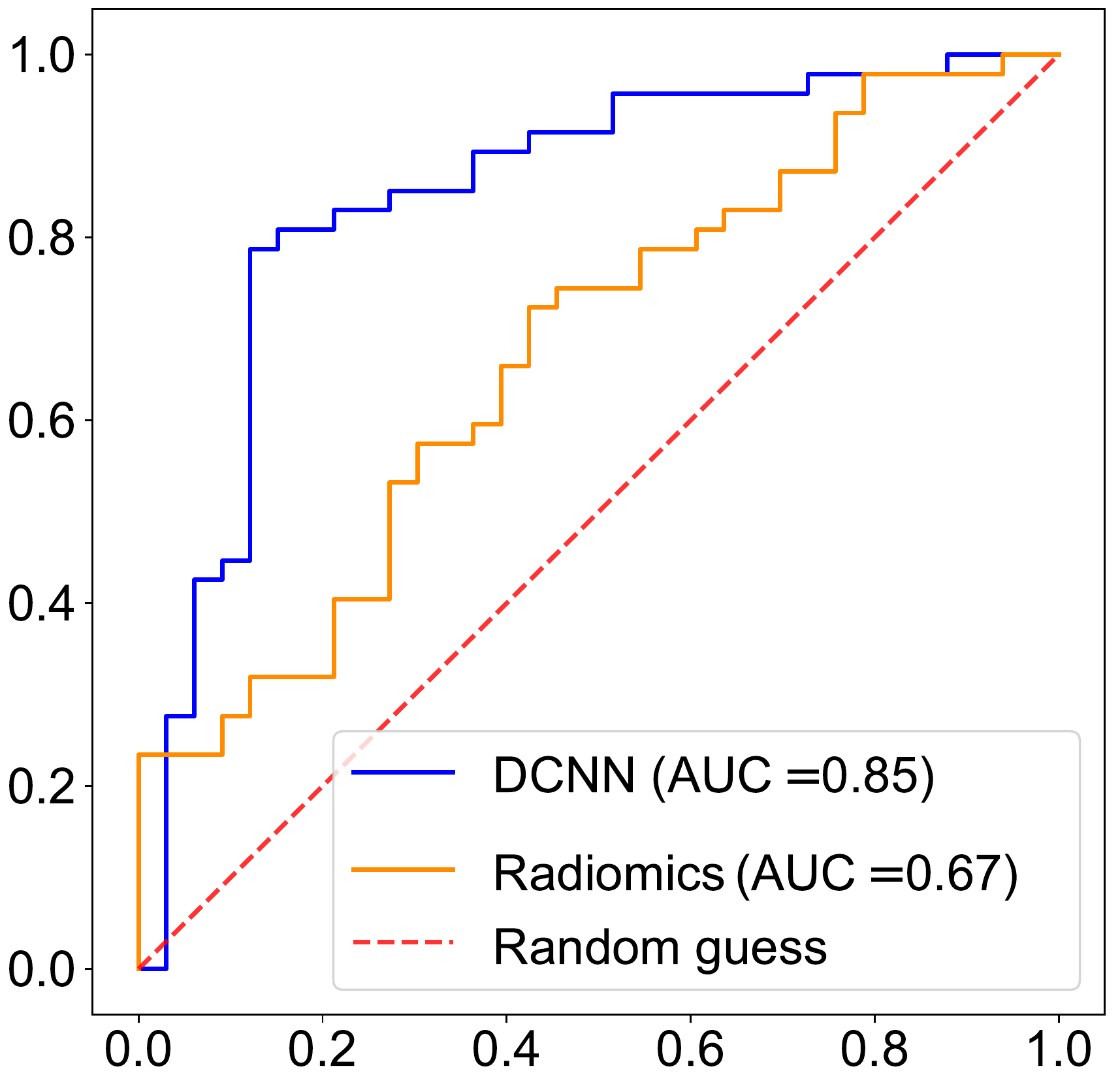}
    }
    \subfloat[IDH mut vs. wt in GBMs]{
        \includegraphics[width=.35\columnwidth]{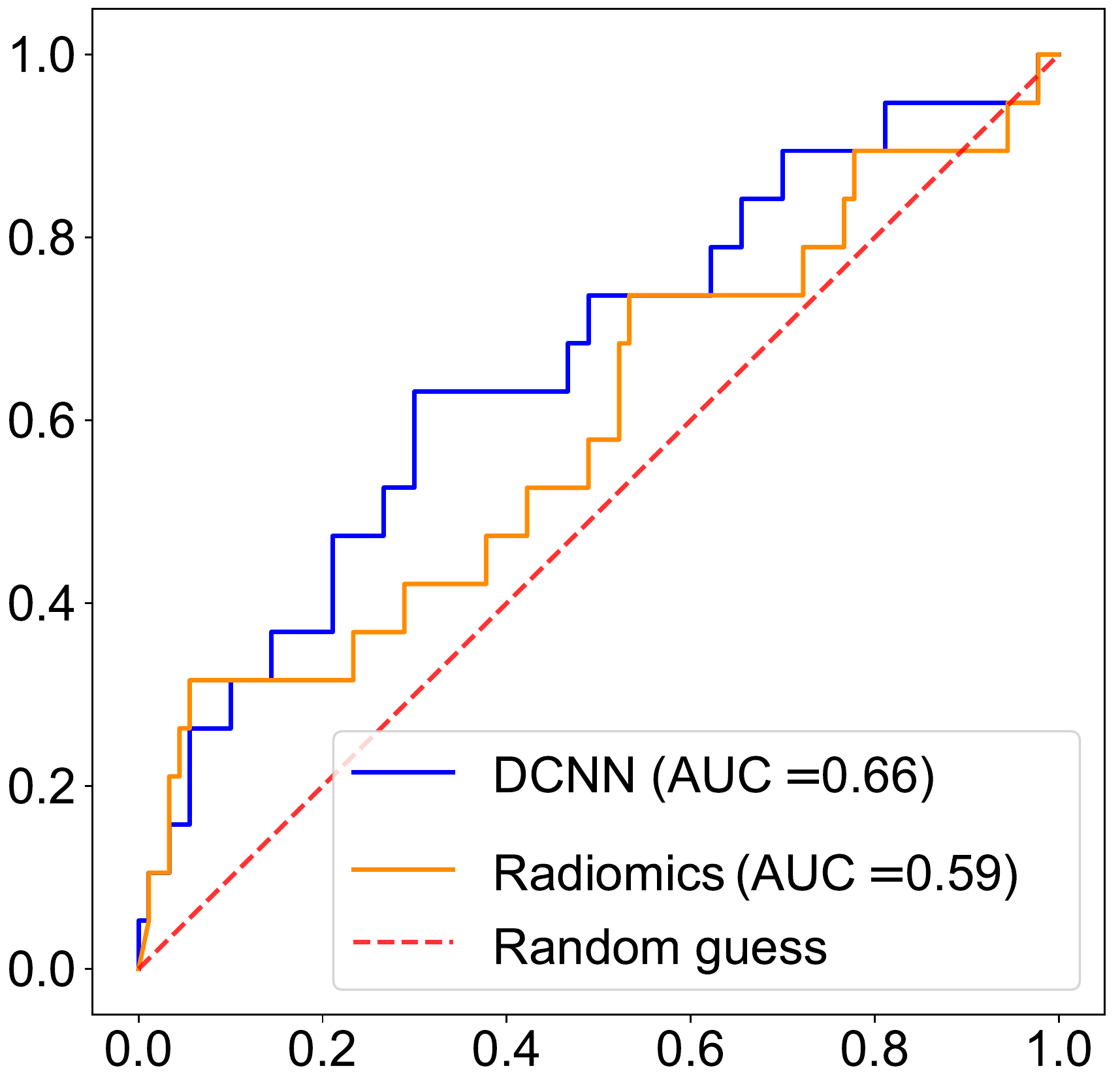}
    }
  \end{minipage}\vspace{9pt}
  \caption{Test ROC curves of our proposed DCNN approach and the radiomics approach proposed in Ref.~\citenum{glioma_sub2018Lu} for the four classification tasks: (a) GBM versus LGG, (b) IDH mut versus wt in LGGs, (c) 1p/19q codel versus noncodel in IDH mut LGGs, and (d) IDH mut versus wt in GBMs.
  The $x$ axes are false positive rates, and $y$ axes are true positive rates.}
  \label{fig:ROCs}
\end{figure}

For evaluation, the AUCs of the ROC curves and accuracies were computed for the trained DCNN models on the test set.
For comparison, the radiomics approach described in Ref.~\citenum{glioma_sub2018Lu} was implemented and evaluated on our data, too.
The ROC curves and experimental results are shown in Fig.~\ref{fig:ROCs} and Table~\ref{tab:test_stats}, respectively.
As can be seen, the DCNN models achieved high accuracies for all the four prediction tasks in the hierarchical classification paradigm, ranging from 0.74--0.83.
The DCNN models also achieved high AUCs for three of the four tasks (0.89, 0.89, and 0.85, respectively).
In addition, in terms of the AUCs, the DCNN models outperformed the radiomics models in all the four tasks (0.66--0.89 versus 0.59--0.85);
and in terms of the accuracies, the former outperformed the latter in three of the four tasks (0.80--0.83 versus 0.64--0.79).
For the fourth task---differentiating between IDH mut and wt in GBMs, neither the DCNN model nor the radiomics model was satisfactory.
The AUCs were 0.66 for the DCNN model versus 0.59 for the radiomics model, and the mean accuracies were 0.64 versus 0.59 (the metric of mean accuracy can better reflect the model performances here than the overall accuracy, given the high inter-class imbalance).
We attribute this phenomenon to the lack of sufficient samples for the subtype IDH mut GBMs, and the corresponding seriously skewed class ratio ($\sim$1:5).
In summary, experimental results suggested the great potential of DCNNs to determine the glioma subtypes in accordance with the WHO 2016 taxonomy, given sufficient, balanced training data.

\begin{table}[ht]
\caption{Model performances on the test set, and a comparison to those of the radiomics approach proposed in Ref.~\citenum{glioma_sub2018Lu}.
For each classification task, the subtype presented first is deemed the positive class, i.e., the GBM, IDH mut, and 1p/19q codel subtypes.
The superior performance in each pair of comparison is bolded.}
\label{tab:test_stats}
\begin{center}
\begin{tabular}{lccccccc}
\hline\hline
Classification (Subject No.) & Model & AUC & Accuracy & Sensitivity & Specificity & Mean Accuracy \\
\hline
GBM vs. LGG & DCNN & \textbf{0.89} & \textbf{0.83} & \textbf{0.81} & \textbf{0.84} & \textbf{0.83}\\
(109 vs. 127) & Radiomics & 0.85 & 0.78 & 0.79 & 0.76 & 0.78\\
\hline
IDH mut vs. wt in LGGs & DCNN & \textbf{0.89} & \textbf{0.80} & \textbf{0.81} & \textbf{0.79} & \textbf{0.80} \\
(80 vs. 47) & Radiomics & 0.75 & 0.69 & 0.69 & 0.68 & 0.68 \\
\hline
1p/19q codel vs. noncodel in & DCNN & \textbf{0.85} & \textbf{0.83} & \textbf{0.85} & \textbf{0.79} & \textbf{0.82} \\
IDH mut LGGs (47 vs. 33) & Radiomics & 0.67 & 0.64 & 0.75 & 0.49 & 0.62 \\\hline
IDH mut vs. wt in GBMs & DCNN & \textbf{0.66} & 0.74 & \textbf{0.47} & 0.80 & \textbf{0.64} \\
(19 vs. 90) & Radiomics & 0.59 & \textbf{0.84} & 0.21 & \textbf{0.97} & 0.59 \\
\hline\hline
\end{tabular}
\end{center}
\end{table}

\section{Conclusions}
\label{sec:conclusion}

In this paper, we presented a DCNN model for noninvasive subtyping of diffuse gliomas with multiparametric MRI data.
As far as the authors are aware of, this is the first study that attempted the comprehensive prediction of all five glioma subtypes as classified by the revised 2016 WHO taxonomy.
The experimental results on a collection of 1,016 real patient data showed that the DCNN model was able to preoperatively predict the molecular subtypes of diffuse gliomas, and that its performances were superior to those of a previously established radiomics approach.
We plan to explore the performance difference between the DCNN and radiomics approaches more deeply by visualization, comparison, and correlation of their respective features.

\acknowledgments 

This study was partially supported by the National Natural Science Foundation of China (No. 81601452), the Key Area Research and Development Program of Guangdong Province, China (No. 2018B010111001), and the Science and Technology Program of Shenzhen, China (No. ZDSYS201802021814180).

\bibliography{report} 
\bibliographystyle{spiebib} 

\end{document}